\documentclass[12pt]{article}
\usepackage{times}
\usepackage[dvips]{graphicx}
\title{Exact Properties of SIQR model for COVID-19}

\author{
Takashi Odagaki$^\ast$\\
\\
\normalsize{Kyushu University}\\
\normalsize{Nishiku, Fukuoka 819-0395, Japan}\\
\normalsize{and}\\
\normalsize{Research Institute for Science Education, Inc.}\\
\normalsize{Kitaku, Kyoto 603-8346, Japan}\\
}

\date{\today}

\begin{document} 

% Double-space the manuscript.
\baselineskip24pt

% Make the title.
\maketitle 

\begin{abstract}
The SIQR model is reformulated where compartments for infected and quarantined
are redefined so as to be appropriate to COVID-19, and exact properties of the
model are presented.
It is shown that the maximum number of infected at large depends strongly on the quarantine rate
and that the quarantine measure is more effective than the lockdown measure in controlling the pandemic.
The peak of the number of quarantined patients is shown to appear some time
later than the time that the number of infected becomes maximum.
On the basis of the expected utility theory, a theoretical framework to find out
an optimum strategy in the space of lockdown measure and quarantine measure is proposed
for minimizing the maximum number of infected and for
controlling the outbreak of pandemic at its early stage.
\end{abstract}

% insert suggested PACS numbers in braces on next line
%\pacs{87.23.Ge,89.65.-s,87.15.A-}

%%%%%%%%%%%%Sect 1
\section{\label{sec:intro}Introduction}

Since November 2019, the pandemic COVID-19 is still expanding in the world, infected of
which totals more than 15 million at 22 nd of July, 2020\cite{JHU}.
It is an important problem of socio-physics to construct a simple model by which one can understand the nature of
the outbreak, and to construct a theoretical framework for formulating the optimum strategy to
control it.\cite{report9}

Epidemics can be considered to be a problem of physics concerning reaction and relaxation
processes and the simplest understanding of its outbreak can be provided by a mean field analysis. 
The SIR model\cite{SIR} assumes three compartments (or species) in population,
susceptible (S), infected (I) and removed (R), and the infection is transferried
from an infected individual to a susceptible individual.
The number of symptomatic patients, which is identical to the number of infected, 
decreases in the community by treatment and/or quarantine.
The SIR model is considered a standard model to explain the infection trajectory of ordinary epidemics
like influenza.

COVID-19 has unusual characteristics: (1) transmission of the virus by presymptomatic
patients and (2) existence of asymptomatic infectious patients,
and (3) patients, symptomatic or asymptomatic, can be identified by PCR test.
Because of these characteristics, the number of infected cannot be obtained directly and
the number of daily confirmed new cases and its time dependence are the only essential observables.
Therefore, COVID-19 showing these characteristics may not be represented properly
by the SIR and the SEIR models which assume that the number of patients is known
and do not treat quarantined patients as a compartment. 

The SIQR model \cite{SIQR,SIQR2009} is a compartmental model which represents a community
by four compartments, assuming an additional compartment of quarantined (Q), 
and describes the transmission process by a system of ordinary
nonlinear differential equations. The SIQR model seems to be appropriate
to COVID-19 and it has been successfully applied in the analysis of the early stage of
the outbreak of COVID-19 in Italy\cite{italy}, India\cite{india,india2}, Sweden\cite{sweden}
and Japan\cite{japan}. 

In this paper, I redefine the SIQR model so as to make it appropriate to COVID-19\cite{japan}.
Namely, I classify patients into two groups; (1) infected patients at large (I)
who can be in any of three states, presymptomatic, symptomatic and asymptomatic and (2)
quarantined patients (Q) who are in a hospital or self-isolated at home and no longer infectious in
the community.
I treat the number of daily confirmed new cases explicitly and consider the quarantine rate or the fraction
of infected at large put in a quarantine or self-isolation as a key parameter
which can be determined from the observation of the daily confirmed new cases.
I also discuss the optimum strategy for controlling the pandemic.

This paper is organized as follows. First, I explain in Sec.~\ref{sec:model} the SIQR model and discuss
its relevancy to COVID-19.
I also present the basic properties of the model, showing parameter dependence of
the maximum number of infected.
Section~\ref{sec:exact} considers the exact solution of the SIQR model and shows the time dependence
of various quantities including the number of quarantined.
A theoretical frame work for optimizing measures to control the outbreak
is discussed in Sec.~\ref{sec:optimum}, where the expected utility theory\cite{EUT}
is exploited. 
The frame work is applied for optimizing strategy for reducing the epidemic peak and
for stamping out the epidemic as fast as possible.
Results are discussed in Sec.~\ref{sec:discuss}.

%%Sect 2
\section{\label{sec:model}SIQR model and basic properties}
\subsection{Model}
The basic concept of the SIQR model is identical to the chemical reaction, which can be described by rate equations.
The dynamics of the SIQR model is given by the following set of ordinary nonlinear
differential equations:
\begin{eqnarray}
\frac{d x}{d t} &=& - \beta x(t) y(t) , \label{dxdt}\\
\frac{d y}{d t} &=&  \beta x(t) y(t) - q y(t) - \gamma y(t) , \label{dydt}\\
\frac{d w}{d t} &=&  q y(t) - \gamma' w(t) , \label{dwdt}\\
\frac{d z}{d t} &=& \gamma y(t) + \gamma' w(t), \label{dzdt}
\end{eqnarray}
where $t$ is the time, and $x(t)$, $y(t)$, $w(t)$ and $z(t)$ are the fractions of population ($N$) 
in each compartment, susceptible, infected at large, quarantined patients and
recovered (and died) patients.
Here, I assumed that new patients immediately after they get infected cannot be quarantined since
the incubation period is long for COVID-19.
The parameters of this model are a transmission coefficient $\beta$,
quarantine rate of infected at large $q$, and
recovery rates $\gamma$ and $\gamma'$ of infected at large and quarantined, respectively.
These parameters can be estimated from observations; $\beta$, $\gamma$ and $\gamma'$ from epidemiological survey
and $q$ from the time dependence of the daily confirmed new cases. Although time can be scaled by one of parametes,
I use one day as a unit of time in this paper
so that ordinary citizens and policy makers can understand results without difficulties.

Infected at large, regardless whether they are symptomatic or asymptomatic,
are quarantined at a per capita rate $q$ and become non-infectious in the community.
Quarantined patients recover at a per capita rate $\gamma '$ (where $1/\gamma '$ is the
average time it takes for recovery) and infected at large become non-infectious at a per capita rate
$\gamma$ (where $1/\gamma$ is the average time that an infected patient at large
is capable of infecting others).
It is apparent that Eqs. (1) $\sim$ (4) guaranteee the conservation of population
$x(t) + y(t) + w(t) + z(t) = 1$.

If one considers quarantined and recoverd together as removed, the set of differential equation
is the same as the set of equations for the SIR model with removal rate of infected $q + \gamma$.
Since the value of $q$ depends strongly on government policies and
the only observable is the number of quarantined patients $\Delta w(t) = q y(t)$ on each day, it is important to
treat the quarantined and recovered patients separately in the analysis of the outbreak of COVID-19.

Figure 1 shows the elementary processes of the SIQR model.

\begin{figure}
\begin{center}
\includegraphics[width=7cm]{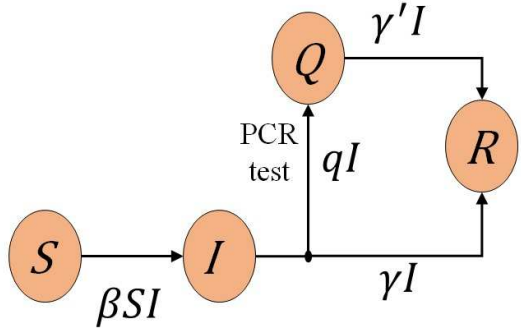}
\caption{Elementary processes of the SIQR model. $x$ and $y$ obey a reaction
$x + y \rightarrow 2y$.}
\end{center}
\label{figure1}
\end{figure}

\subsection{Basic properties}
From Eqs. (1) and (2), it is easy to show that the trajectory in the $(x, y)$ plane is
determined by
\begin{equation}
\frac{dy}{dx} = \frac{q + \gamma}{\beta x} - 1 .
\label{dydx}
\end{equation}
Therefore, the trajectory in the $(x, y)$ plane is given by
\begin{equation}
y = 1 - x + \frac{q + \gamma}{\beta} \ln x ,
\label{y-xtrajectory}
\end{equation}
where the initial condition is set to $y = 0$ at $x = 1$.
Figure 2 shows the trajectories (a) for various $\beta$ at $q=0.1$ and (b) for various $q$ at $\beta = 0.4$
when $\gamma = 0.06$.

\begin{figure}%%figure 2
\begin{center}
\includegraphics[width=6cm]{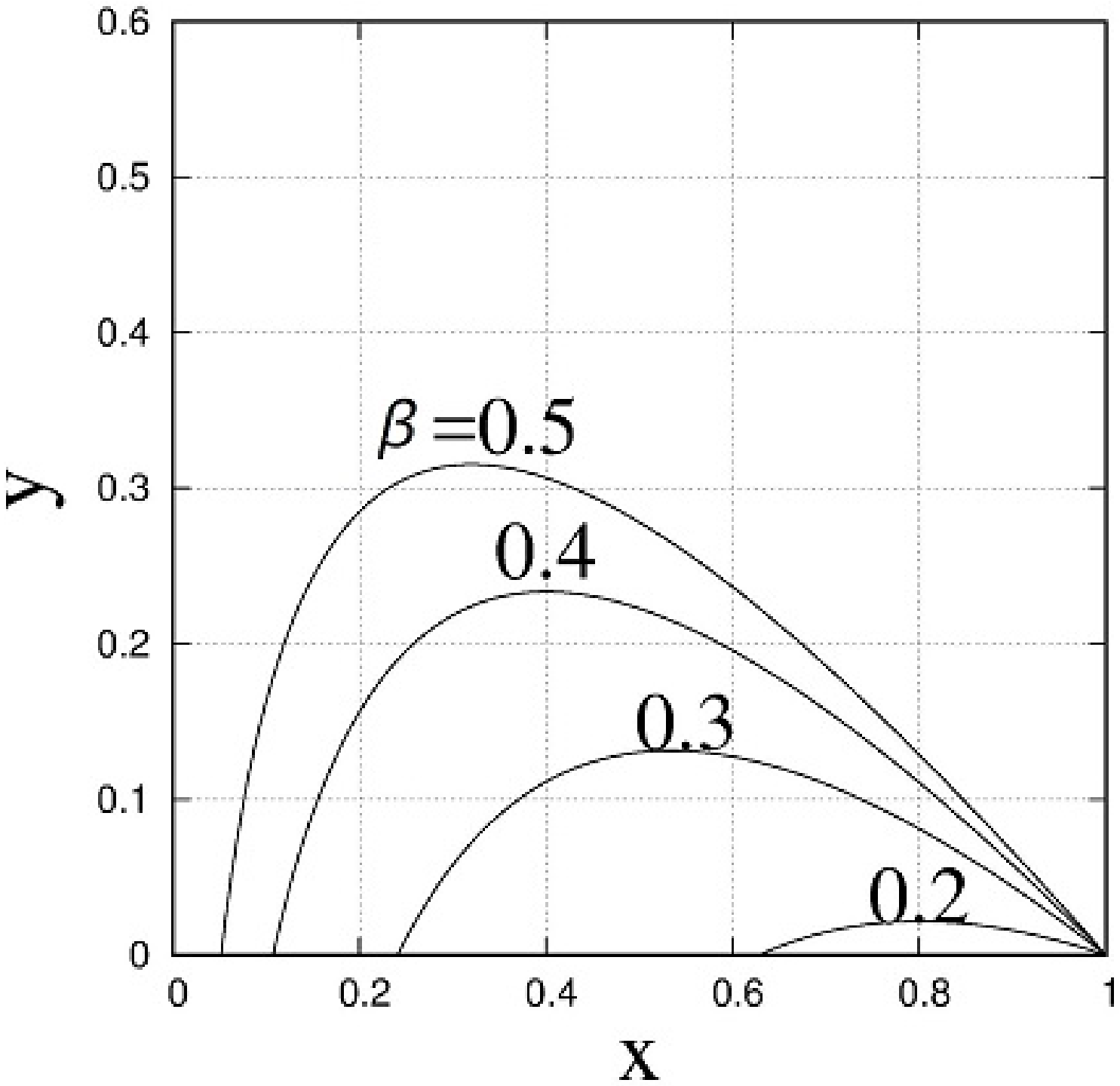}
\includegraphics[width=7.5cm]{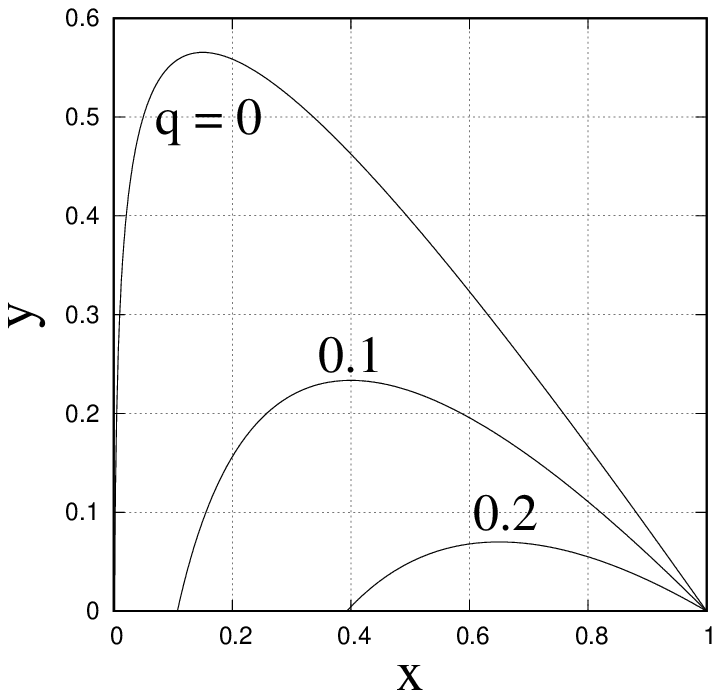}\\
(a) \hspace{7cm}(b)
\caption{Trajectories in the $(x, y)$ plane.
(a) for various $\beta$ when $q=0.1$ and (b) for various $q$ when $\beta = 0.4$,
where $\gamma = 0.06$.}
\end{center}
\label{figure2}
\end{figure}

The peak position $ (x^*, y^*)$ of the trajectory can be obtained from Eqs. (\ref{dydx}) and (\ref{y-xtrajectory})
by setting $\frac{dy}{dx} = 0$. I find that
\begin{eqnarray}
x^* &=& \frac{q + \gamma}{\beta}, \\
y^* &=& 1 - \frac{q + \gamma}{\beta} + \frac{q + \gamma}{\beta} \ln \frac{q + \gamma}{\beta} .
\label{ypeak}
\end{eqnarray}

Figure 3 shows the dependence of $y^*$ on $\beta$ and $q$:
(a) three dimensional plot of $y^*(\beta, q)$, (b) the $\beta$ dependence at $q = 0.1$
and (c) the $q$ dependence at $\beta = 0.4$ when $\gamma = 0.06$.
The peak height becomes lower for smaller $\beta$ and enhanced $q$,
and the latter is more effective in reducing the peak.

\begin{figure}
\begin{center}
\includegraphics[width=12cm]{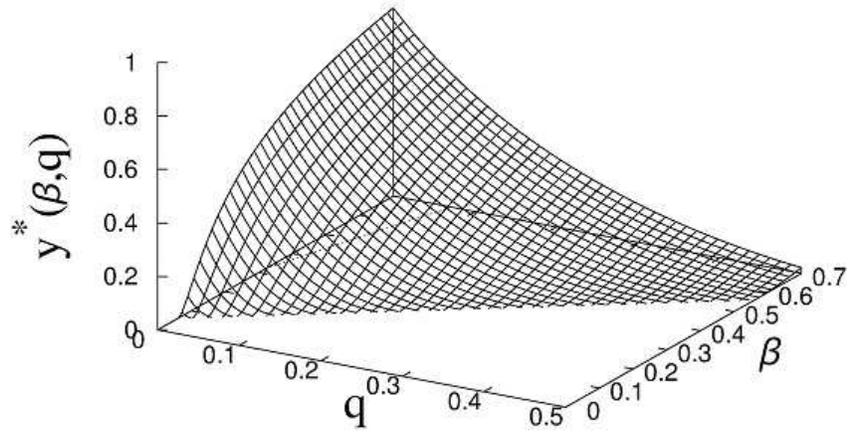}\\
(a)\\
\includegraphics[width=7cm]{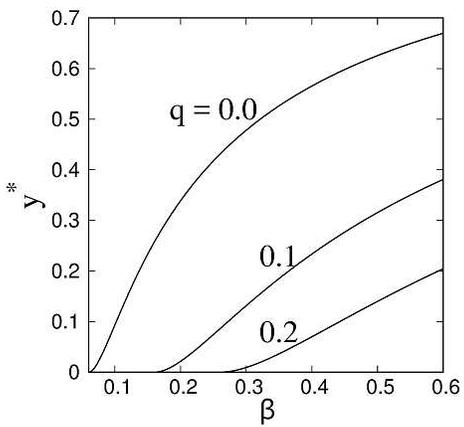}\hspace{0.5cm}
\includegraphics[width=7cm]{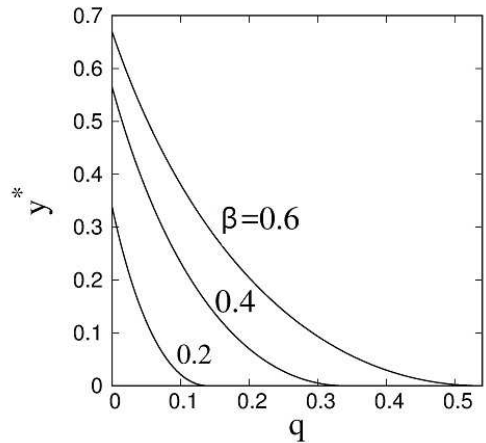}\\
(b) \hspace{7cm} (c)
\caption{Peak heights of infected at large as a function of $\beta$ and $q$.
(a) Three dimensional plot, 
(b) for various $q$ when $\beta = 0.4$ and
(c) for various $\beta$ when $q = 0.1$, where $\gamma = 0.06$. }
\end{center}
\label{figure3}
\end{figure}

%%Sect 3
\section{\label{sec:exact}Time dependence of the outbreak}
\subsection{Exact properties}
Combining Eqs. (\ref{dwdt}) and (\ref{dzdt}) together, I obtain
\begin{equation}
\frac{d (w + z)}{dt} = (q + \gamma) y(t)
\label{dwzdt}.
\end{equation}
Therefore, the set of Eqs. (\ref{dxdt}), (\ref{dydt}) and (\ref{dwzdt}) is
identical to the basic equations of the SIR model as stated before,
and the exact solution can be written as\cite{SIRexact}
\begin{eqnarray}
x(t) &=& x_0 u(t), \\
y(t) &=& 1 - x_0 u(t) + \frac{q + \gamma}{\beta} \ln u(t), \\
w(t)+z(t) &=& -\frac{q + \gamma}{\beta} \ln u(t),
\end{eqnarray}
where time $t$ is related to $u$ through an integral
\begin{equation}
t = \int_1^u \frac{d\xi}{\xi [ x_ 0\beta \xi - \beta - (q + \gamma)\ln \xi]} .
\label{tandu}
\end{equation}
Once $y(t)$ is known, then $w(t)$ is obtained from Eq. (\ref{dwdt}),
\begin{equation}
w(t) = q e^{-\gamma' t}\int_0^t e^{\gamma' t'} y(t') dt' ,
\end{equation}
and thus $z(t)$ is given by
\begin{equation}
z(t) =  -\frac{q + \gamma}{\beta} \ln u(t) - q e^{-\gamma' t}\int_0^t e^{\gamma' t'} y(t') dt'.
\end{equation}

Figure 4 shows the time dependence of $x(t)$, $y(t)$, $w(t)$ and $z(t)$
for $\beta = 0.4$, $q = 0.1$ and $\gamma = \gamma' = 0.06$.
It is interesting to note that the number of quarantined patients $w(t)$ takes its maximum
at some time later than the time that the number of infected $y(t)$ becomes maximum.
\begin{figure}
\begin{center}
\includegraphics[width=8cm]{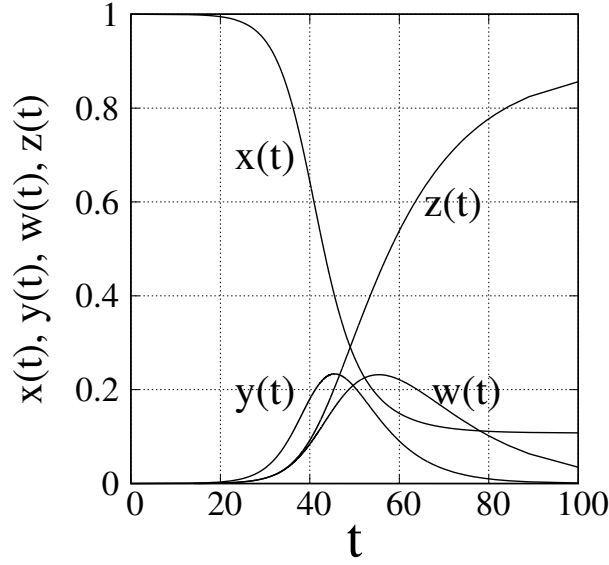}\\
\caption{The exact time dependence of $x(t)$, $y(t)$, $w(t)$ and $z(t)$
when $\beta = 0.4$, $q = 0.1$ and $\gamma = \gamma' = 0.06$. 
The peak positions of $y(t)$ and $w(t)$ differ about 10 days for the choice of parameters.}
\end{center}
\label{figure4}
\end{figure}

The observable in COVID-19 is the daily confirmed new cases $\Delta w(t) = q y(t)$, which is simply
$q$-times smaller than $y(t)$ as shown in Fig. 5.
\begin{figure}
\begin{center}
\includegraphics[width=8cm]{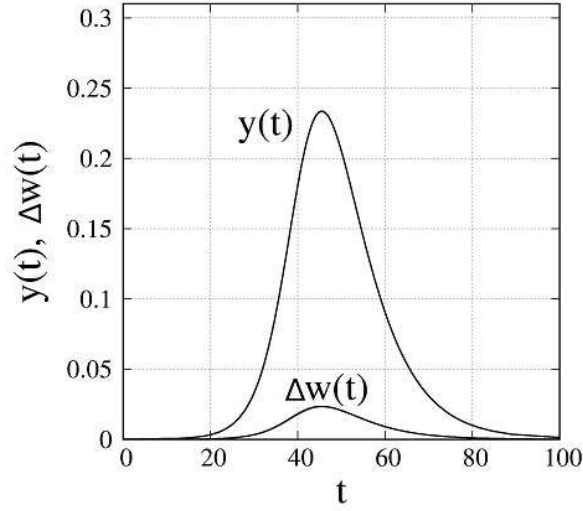}\\
\caption{The time dependence of $y(t)$ and $\Delta w(t)$
for $\beta = 0.4$, $q = 0.1$ and $\gamma = \gamma' = 0.06$. }
\end{center}
\label{figure5}
\end{figure}

\subsection{Properties at the early stage of outbreak}
Since $x(t) \simeq x_0$ and $y_0 \simeq 0$ in the early stage of outbreak,
I introduce a new integration variable $\eta = 1- \xi$ in Eq. (\ref{tandu}) 
\begin{equation}
t = - \int_0^{1-u} \frac{d\eta}{(1-\eta) [ x_ 0\beta (1-\eta) - \beta - (q + \gamma)\ln (1-\eta)]} .
\label{tandu2}
\end{equation}
Taking up to the first order term in $1-u$ in the integrand on the right-hand side
of Eq. (\ref{tandu2}) and up to the first order term in $y_0$, I obtain
\begin{equation}
u(t) = 1 + \frac{y_0 \beta}{q + \gamma - \beta}\left( e^{(\beta - q - \gamma)t} - 1 \right) .
\label{uoft}
\end{equation}
Therefore, the short-term solution of the SIQR model is given by
\begin{eqnarray}
x(t) &=& x_0 - \frac{y_0\beta}{\beta-q - \gamma}\left(e^{(\beta - q - \gamma)t} - 1 \right), \\
y(t) &=& y_0 e^{(\beta - q - \gamma)t}, \\
w(t) &=& \frac{y_0 q}{\beta - q - \gamma + \gamma'}\left( e^{(\beta - q - \gamma)t} - e^{-\gamma' t} \right),\\
z(t) &=& \frac{y_0 (q + \gamma)}{\beta - q - \gamma}\left( e^{(\beta - q - \gamma)t} -1 \right) -
\frac{y_0 q}{\beta - q - \gamma + \gamma'}\left( e^{(\beta - q - \gamma)t} - e^{-\gamma' t} \right) .
\end{eqnarray}
These solution can also be obtained by setting $x(t) = x_0 = 1$ in Eqs. (\ref{dxdt}) $\sim$ (\ref{dzdt})
and have been used in the analysis of the outbreak of COVID-19 \cite{italy,india,india2,japan}.

The short-term solution indicates that the initial growth rate of the number of infected at large
is determined by 
\begin{equation}
\lambda = \beta - q -\gamma.
\label{growthrate}
\end{equation}
In order to control the outbreak, a measure must be formulated to make the growth rate $\lambda$
negative under various restrictions in economic activities and medical care systems.

%%Sect 4
\section{\label{sec:optimum}Optimum measure}
\subsection{General frame work}
As shown in Fig.~4, the present model like the SIR model shows that the epidemic curve will converge to an equilibrium state
after several months, passing a maximum number of infected which could be $15 \sim 25$ \% of population
depending on the value of parameters.
This means that if we wait the natural epidemic equilibrium, the number of causalities will become
unacceptably large.
In order to reduce the number of causalities, all governments in the world have been struggling against COVID-19
with various measures. Lockdown and social distancing are measures to reduce the effective
transmission coefficient $\beta$, but it has a severe damage on economic activities.
For COVID-19, quarantine measure including self-isolation has been employed in many countries
to increase the quarantine rate $q$.

Since $\gamma$ in the SIQR model cannot be altered, $\beta$ and $q$ are the essential parameters
which can be modified by policy. I parameterize $\beta$ as $\beta = (1-a)\beta_0$ where
$a$ represents the strength of lockdown measure; $a=0$ corresponds to no measure on social distancing
and $a=1$ denotes complete lockdown. I consider a measure ${\cal M} = {\cal M}(a,q)$ which is a function of
$a$ and $q$.
The transmission rate $\beta_0$ can be determined from the growth rate of the epidemic at the earliest stage
when no measures are imposed.

I consider a cost function ${\cal C}(a, q)$ of a measure characterized by $a$ and $q$ which must be
an increasing function of  $a$ and $q$. The problem is to move ${\cal  M}$ in a desired direction,
making the cost as small as possible. Namely, for a given ${\cal M}$, an optimum set $(a, q)$
minimizing the cost is obtained which in turn determines the optimum trajectory in the $(a, q)$
space.
I introduce a Lagrange multiplier $\mu$ and consider a function  ${\cal F}(\mu, a, q)$ defined by
\begin{equation}
{\cal F}(\mu, a, q) = {\cal C}(a, q) - \mu \left[ {\cal M} - {\cal M}(a, q) \right].
\end{equation}
It is well known that the optimum value of  $C(a, q)$ is given by the solution to
\begin{eqnarray}
\frac{\partial {\cal F}}{\partial a} &=& 0 ,\\
\frac{\partial {\cal F}}{\partial q} &=& 0 ,\\
\frac{\partial {\cal F}}{\partial \mu} &=& 0 .
\end{eqnarray}

In the following discussion, I consider a model cost function in an arbitrary unit\cite{Yan}
\begin{equation}
{\cal C}(a, q) = a^2 + k \left( \frac{q}{\beta_0}\right)^2.
\label{cost}
\end{equation}
Here, $k$ is a parameter characterizing relative importance of measures. When $k = 1$ , both measures cost equally.
When $k > 1$, cost for medical treatment is larger than the economic cost, and when $k < 1$, economic cost
due to social distancing is larger than the medical cost.

\subsection{Optimum measure to reduce the epidemic peak}
As shown in Fig.~3, the epidemic peak and hence the number of quarantined strongly depend
on parameter $a$ and $q$.
Taking Eq. (\ref{ypeak}) as a measure, I set
\begin{equation}
{\cal M}(a, q) = 1 - \frac{q + \gamma}{(1-a)\beta_0} + \frac{q + \gamma}{(1-a)\beta_0} \ln \frac{q +
\gamma}{(1-a)\beta_0}.
\end{equation}
It is straightforward to find that the optimum trajectory $(a^*, q^*)$ satisfies
\begin{equation}
a^*(1-a^*)\beta_0^2 = kq^*(q^* + \gamma).
\end{equation}
Figure 6 shows the optimum strategy for $k = 2, 1, 0.5$.
\begin{figure}
\begin{center}
(a) \includegraphics[width=6.5cm]{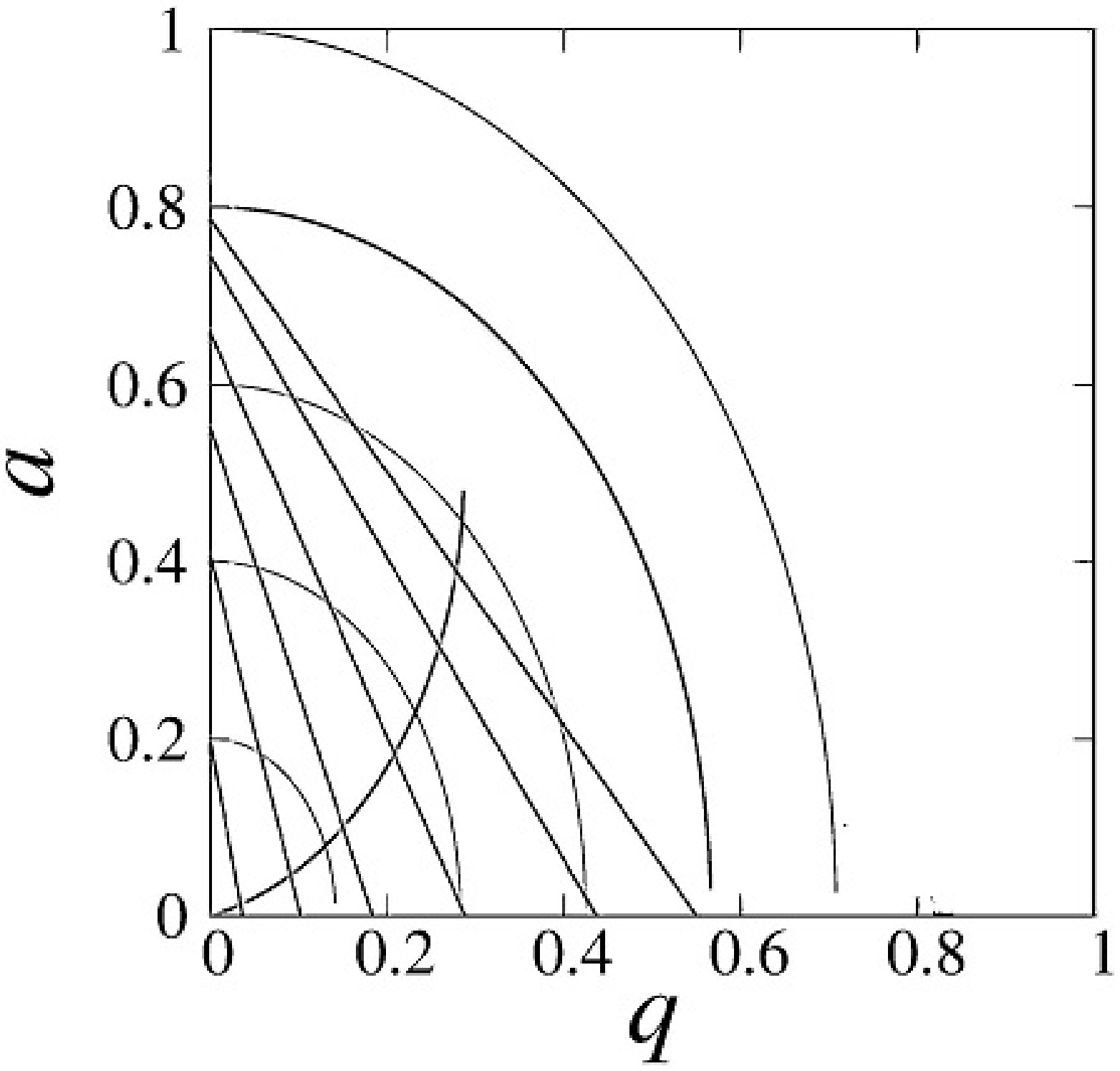} \\
(b) \includegraphics[width=6.5cm]{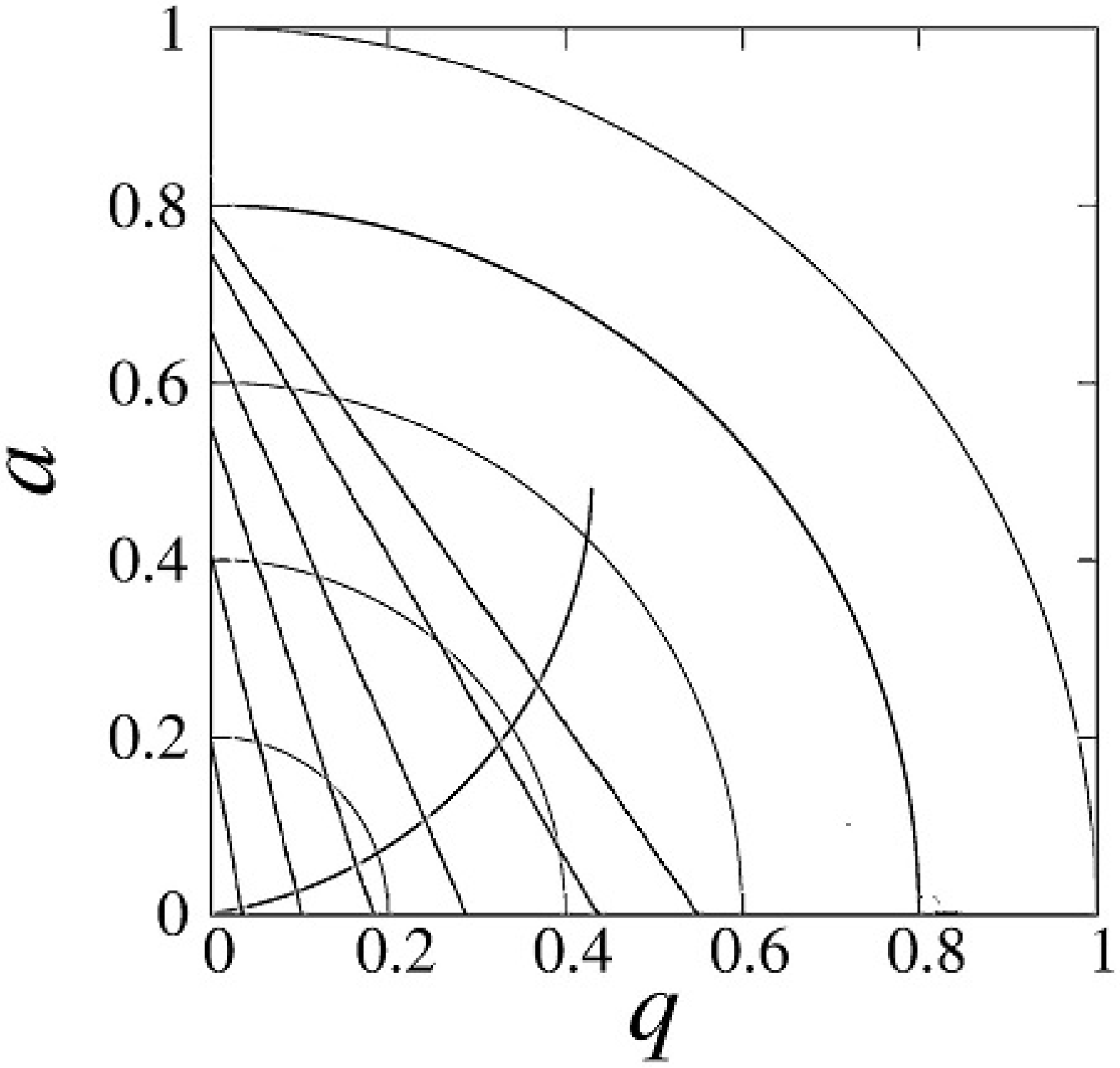} \\
(c) \includegraphics[width=6.5cm]{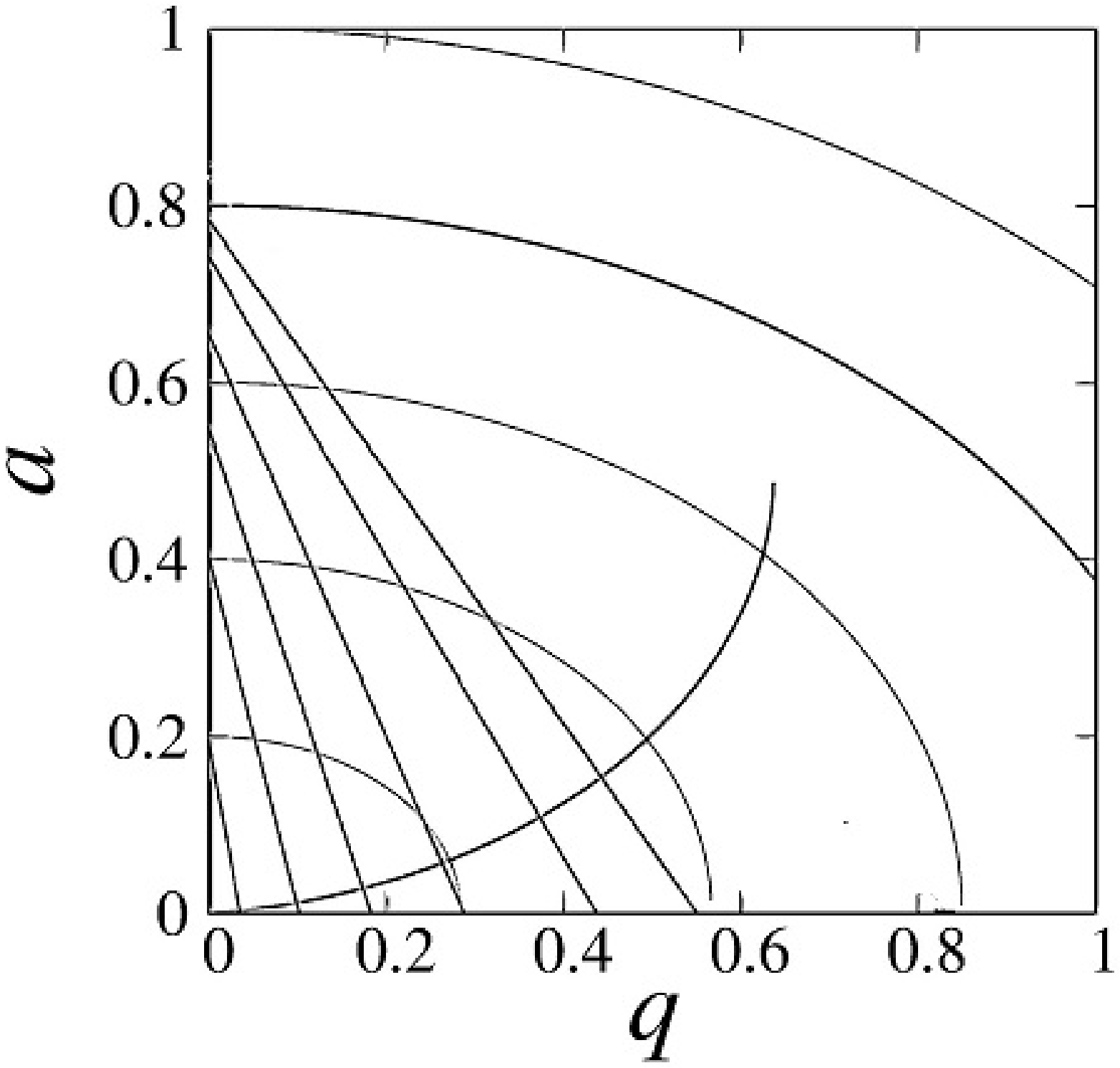}  \\
\caption{The solid thick curve shows the trajectory of the optimum strategy to reduce the epidemic peak
for $\gamma =  0.06$. Straight lines are constant $y^{*}$ lines: $y^{*} = 0.5, 0.4, 0.3, 0.2, 0.1, 0.05$ from left to right,
and oval curves represent constant cost curves: $C = 0.04, 0.16, 0.36, 0.64, 1$ from left to right.
(a) $k = 2$, (b) $k = 1$ and (c) $k = 0.5$.
} 
\end{center}
\label{figure6}
\end{figure}

\subsection{Optimum measure to stamp out the epidemic}
In this subsection, I discuss an optimum measure to stamp out the epidemic
at the beginning of the outbreak.
The growth rate of the number of infected at large is given by Eq.~(\ref{growthrate})
in the early stage of the outbreak, and I set 
\begin{equation}
{\cal M}(a,q) = (1 - a)\beta_0  - q - \gamma.
\end{equation}
The aim of a measure is to bring the state from the region ${\cal M}(a,q) > 0$ at $(a = 0, q=0)$
to some region ${\cal M}(a, q) < 0$.

It is straightforward to find that the optimum trajectory $(a^*, q^*)$ satisfies
\begin{equation}
a^* = k\frac{q^*}{\beta_0}.
\end{equation}

Figure 7 shows the optimum trajectory on the $(a, q)$ plane: (a) $k = 2$,
(b) $k = 1$ and (c) $k = 0.5$.

\begin{figure}
\begin{center}
(a) \includegraphics[width=6.5cm]{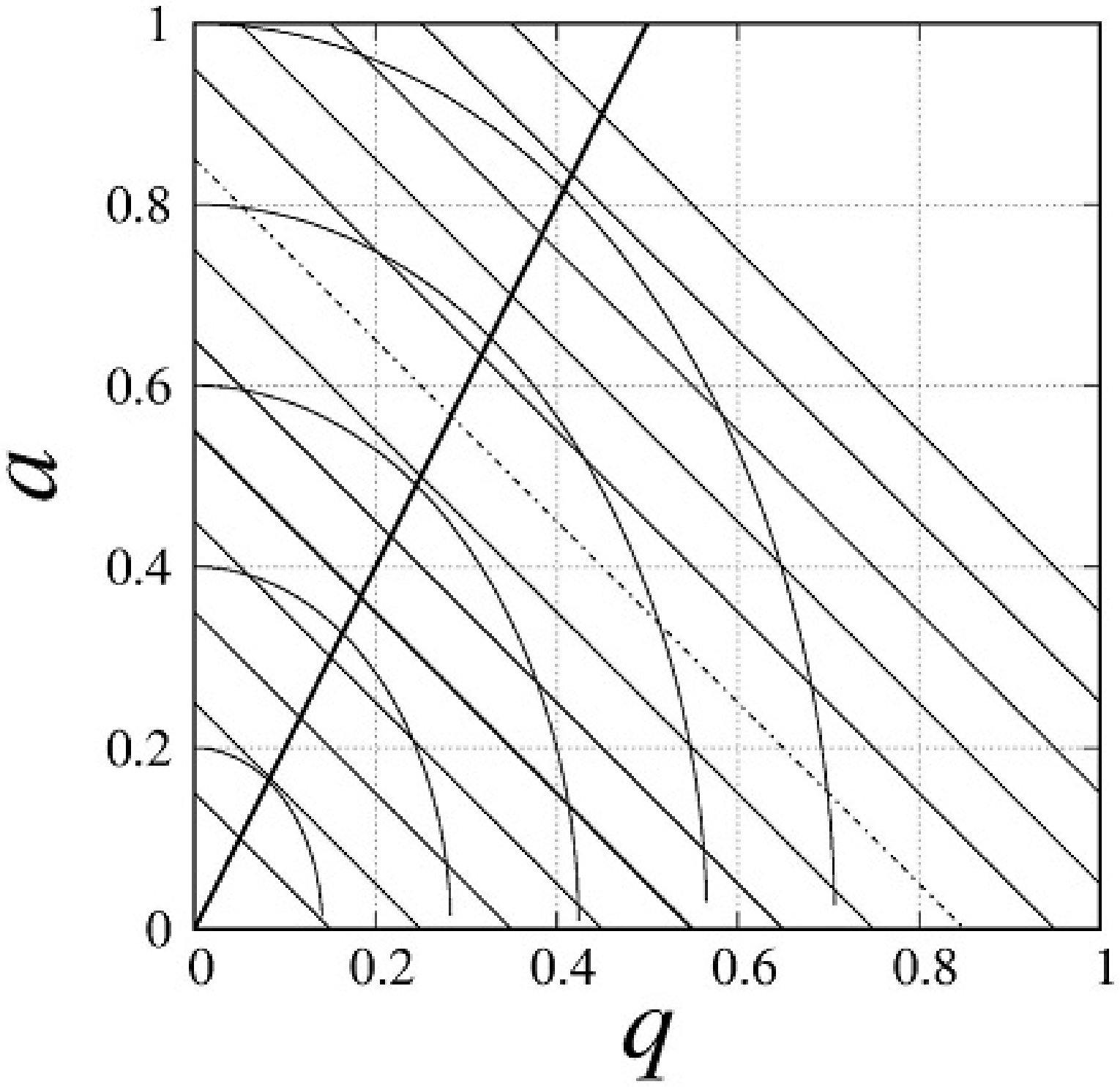} \\
(b) \includegraphics[width=6.5cm]{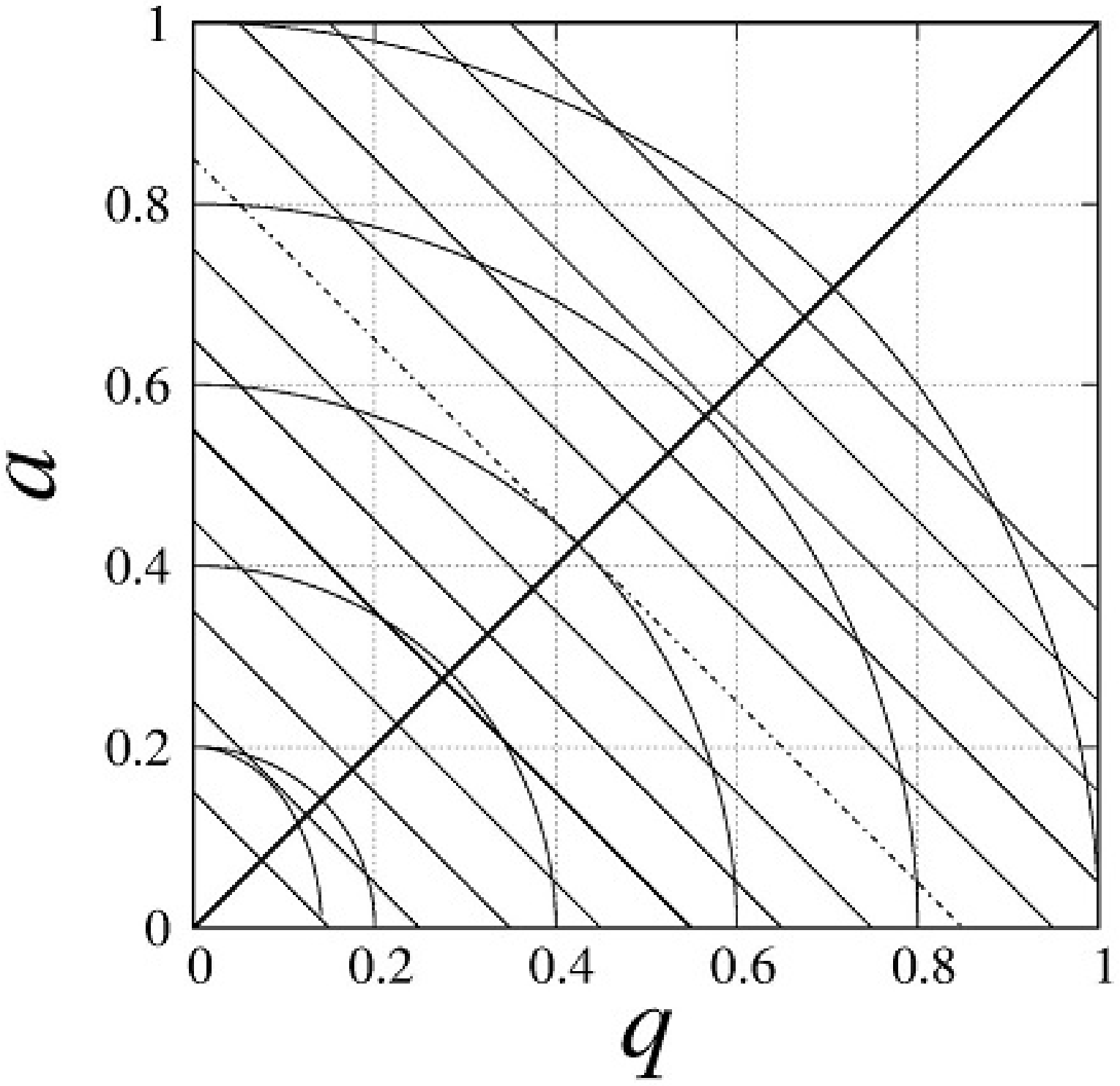} \\
(c) \includegraphics[width=6.5cm]{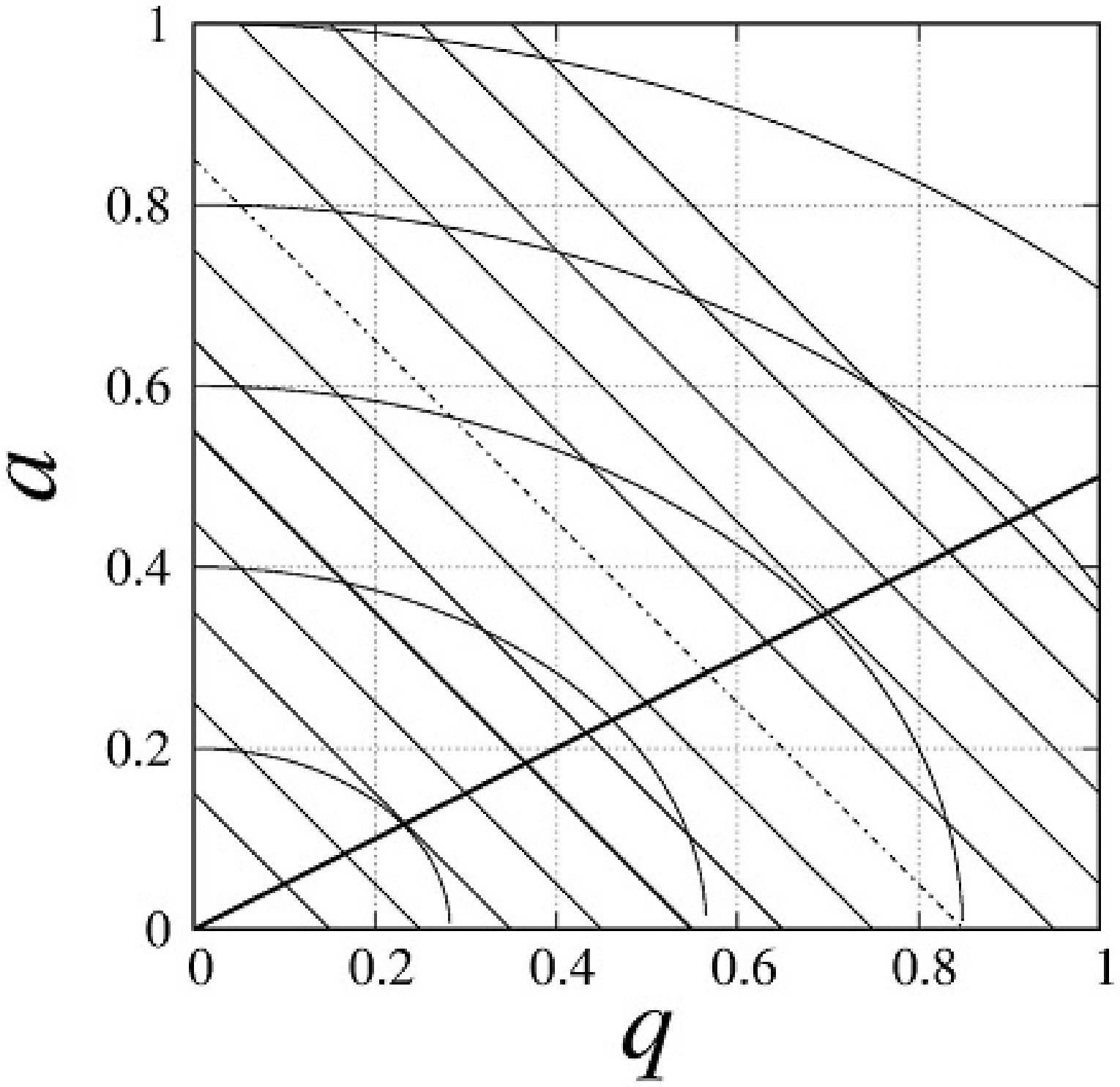}
\caption{The increasing thick straight line is the trajectory of the optimum strategy to stamp out the epidemic
for $\gamma =  0.06$. 
Decreasing straight lines shows the constant growth rate lines from 0.7 (the lowest line)to -0.4 (the highest line),
and the dashed line shows $\lambda = 0$. Oval curves represent constant cost curves: $C = 0.04, 0.16, 0.36, 0.64, 1$
 from left to right.
(a) $k =2$, (b) $k = 1$ and (c) $k = 0.5$.
} 
\end{center}
\label{figure7}
\end{figure}

%%%%Sect 6
\section{\label{sec:discuss}Discussion}
In this paper, I presented exact properties of the SIQR model relevant to COVID-19 in the entire
time span and in the early stage of the outbreak.
In particular, I investigated dependence of the outbreak on parameters, transmission coefficient and quarantine
rate, which can be controlled by measures.
It is important to note that the peak of the number of quarantined patients $w(t)$ appears 
about 10 days later than the time that the number of infected $y(t)$ becomes maximum,
for the choice of parameters used in Fig.~4.
Within this model, the number of infected at large $y(t)$ can be estimated from 
the daily confirmed new cases $\Delta w(t)$, once the quarantine rate $q$ is obtained 
from the infection trajectory. 
I also discussed a theoretical framework for optimizing measures to control the outbreak
on the basis of the expected utility theory, when the social cost is given by 
a function of the social distancing policy and the quarantine measure.
The optimum strategy depends on the aim; reducing the epidemic peak or accelerating the stamping out the
outbreak.
It should be emphasized that a simple lockdown ($a = 1$) is not the optimum strategy.

Given the cost function and the aim of policy in each country,
it will be possible to formulate the optimum policy specific to the country for controlling the outbreak
on the basis of the present theoretical framework.

The SIQR model does not consider any memory effects of the epidemic like the incubation period
and the infectious period functions.
It is an important future problem to include memory effects of $\beta$ and $q$ in the SIQR model.

%\section*{Acknowledgments}
%I would like to thank Drs. M. Matsushita, M. Sano, Y. Yamazaki and R. Fujie
%for valuable discussion. My deepest appreciation goes to my family
%who helped me in preparation of this manuscript.

\end{document}